\documentclass{elsart}
\usepackage{natbib}
\usepackage{graphicx}
\usepackage{amssymb}

\begin{document}

\begin{frontmatter}


\title{Low-mass bare strange stars}

\author{R. X. Xu}
\address{School of Physics, Peking University, Beijing, China 100871}

\begin{abstract}
\label{abs}

Strange stars with low masses are suggested to exist in reality,
the origin of which could be via accretion-induced collapse (AIC)
of white dwarfs. Such a strange star is likely bare, and would
thus spin very fast, even to a period of $<0.1$ ms. Strange stars
with low masses may differ from those with solar-masses in various
astrophysical appearances. Observations to test this ``low-mass''
idea are proposed.

\end{abstract}


\end{frontmatter}

{\bf 1. Introduction}

The existence of quark matter is a direct consequence of the
asymptotically free nature of the strong interaction, which was
proved in 1973 based on non-Abelian gauge theories. How can we
find quark matter in reality? Besides the exploration in
relativistic heavy ion colliders, strange (quark) stars, which are
composed of almost equal numbers of $u$, $d$, and $s$ quarks, are
likely the stable bulk quark matter expected to be born after
supernova explosions. The knowledge of elementary
color-interaction could be improved greatly via studying pulsars
if they are strange stars, rather than normal neutron stars.

Pulsar-like stars could have very low masses in strange star
models, because the mass of strange quark matter can be as low as
a few hundreds of baryons due to color confinement, thought the
maximum mass can be order of $M_\odot$.
Low-mass bare strange stars and relevant issues are studied in
\cite{xu04}, including the period evolution in both rotation- and
accretion-powered phases, the proposal that pulsar strong magnetic
fields are ferromagnetism-originated, the suggestion that the
quark surfaces could be essential for successful supernova
explosions of both solar- and low-mass strange stars, some
candidates with low-masses, and possible observational tests.
The low-mass strange stars are to be studied further in this
paper, with attention being paid to their astrophysical
appearances in order to identify one.

{\bf 2. Possible origin of low-mass strange stars}

Fermion stars stand against gravity by degenerate pressures, but
are failed to do so if their masses, $M$, are higher than the mass
limits, $M_{\rm ch}$. Electron-degenerated matter (e.g., white dwarfs) should collapse,\footnote{%
Another possibility is an unstable thermal nuclear burst when
$M\sim M_{\rm ch}$, which appears as a Type Ia supernova
explosion.
} %
and the gravitational (and maybe others) energy release may result
in mass-ejection, when $M>M_{\rm ch}$.
After collapse, neutron-rich hadron matter may emergence
temporarily, and then the degenerate pressure of a deeper layer of
constituents below neutrons, the quarks, could counter-balance
gravity if the density is high enough to de-confine quarks.
A quark star is thus form, and the star could be bare and with
strangeness if the Witten's conjecture, that bulk strange quark
matter is absolutely stable, is correct.

Two scenarios for the collapse of electron-degenerated matter. 1,
the iron-cores in the center of evolved massive stars are
degenerate, the masses of which increase as nuclear fusion
processes, and the cores therefore have to collapse if $M>M_{\rm
ch}$. 2, the mass of an accreting (from its binary or just of
fall-back interstellar-medium) white dwarf increases, and will
collapse finally.
The former is one of the conventional mechanisms of supernova
explosions, and the later is the so-called accretion-induced
collapse (AIC) of white dwarfs (of CO or O-Ne-Mg). Whether an AIC
process can result in the formation of a neutron stars (or a quark
star) is still a matter of debate both observationally and
theoretically.
Nevertheless, if both scenarios can produce quark stars, the
properties of residual quark stars may not be the same, since one
scenario differs significantly from the other.
\cite{xu04} suggests that quark stars with low masses may form in
AIC processes, while iron-core collapses could generally create
normal-mass quark stars.

Actually, there are some hints of AIC-produced neutron stars.
A general review on the possible connection between white dwarfs
and neutron stars can be find in \cite{cg97}, and a short review
on applying the AIC-idea to explain various observations in
\cite{f99}.
The existence of pulsar-like stars with low-mass ($\sim M_\odot$)
companion raises a question: how might such systems have survived
to core-collapse supernovae without being disrupted by strong
mass-ejection (e.g, a 20$M_\odot$-progenitor will eject $\sim
18M_\odot$ material after explosion)?
%
%
It is suggestive that millisecond pulsars are born via AIC of
white dwarfs in order to avoid those troublesome issues, but
assuming currently that the mass-ejection is small during AIC.
This assumption may not be reasonable in view of the fact that
most of the progenitor mass would be ejected during an iron-core
collapse. If mass-ejections are similar in these two cases, one
have to conclude that AIC-originated pulsars could be low massive.

Low-mass millisecond pulsar idea could not be so good in the
normal neutron star model, but sounds reasonable in the strange
star model.
Low-mass neutron stars have large radii, which could not rotate
very fast. Note: the Kepler period of the minimal mass neutron
star could be (mass $\sim 0.1M_\odot$ and radius $\sim 160$ km) $
%
P_{\rm k}=2\pi/\sqrt{GM/R^3}\simeq 110
$ %
ms, which is much larger than the period of millisecond pulsars (1
ms $\sim$ 50 ms).
However, low-mass bare strange stars have an advantage of fast
rotation \citep{xu04}.
In this sense, simulating the AIC process (e.g., to know the
mass-ejection) could be helpful for identifying quark stars.
In adition, a rapid rotating star in the AGB-phase may probably
have enough mass to collapse when the star spins down.
We may expect that a great percentage of millisecond pulsars
(through a channel of $(2\sim 8)M_\odot$-{\rm star} $\rightarrow$
{\rm white dwarf} $\rightarrow AIC\rightarrow$ {\em low-mass
strange star}) exist in the Galaxy due to the high population of
stars with low and intermidiate masses.

It is not clear whether the very little timing noise observed in
millisecond pulsars relates to the low masses of such pulsars.
Intuitively, an active (quiet) quantity varies more (less)
significantly. However the spin frequency, $\Omega$, of pulsars
does not obey this rule: the timing noise is high for low $\Omega$
(normal pulsars), but is low for high $\Omega$ (millisecond
pulsars).
Although a few models are invented to understand this unusual
issue, a simple suggestion could be that those two kinds of
pulsars have different masses, since an object (e.g., a
millisecond pulsar) with low mass should be less active and thus
stable.

{\bf 3. The conditions for keeping strange stars bare}

It is worth knowing whether {\em bare} strange stars are
ubiquitous, since the exposed quark surfaces are very important
for us to identify strange stars and to receive information
directly from quark matter.
A newborn strange star should be bare due to detonating
combustion, but it could be crusted in its later history with high
accretion rate \citep{xu02}.

In case of low-mass bare strange stars, two conditions should be
satisfied in order for keeping them bare. (1), A single accreted
ion (e.g., a proton) should have enough kinematic energy to
penetrate the Coulomb barrier: $GMm_p/R>\sim V_q$. The Coulomb
barrier, $V_q$, is model-dependent, which varies from $\sim 20$
MeV to even $\sim 0.2$ MeV (in cases of low strange quark mass and
high color coupling). Approximating $M=(4/3)\pi R^3(4B)$ ($B$ is
the bag constant), one has
\begin{equation}
M>\sim \sqrt{3V_q^3\over 16\pi B G^3 m_p^3} \simeq 6.5\times
10^{-4} V_{q1}^{3/2} B_{60}^{1/2} M_\odot
\end{equation}
from the first condition, where $V_q=V_{q1}$ MeV, $B=B_{60}\times
60$ MeV fm$^{-3}$.
(2), The accreting material is not halted significantly even near
stellar surfaces. The critical accretion luminosity, beyond which
crusts form, is \citep{xu02},
%
$L^* = {32\sqrt{2}\pi^{5/2}\sqrt{c}Gm_pB\over 3\epsilon
\sigma_T}{R^{7/2}\over P^{1/2}}\simeq 8\times
10^{33}{R_1^{7/2}\over \epsilon P_{\rm ms}^{1/2}}~{\rm erg/s}$,
%
%
where radius $R=R_1$ km, spin period $P=P_{\rm ms}$ ms, and the
effect, that only $\varepsilon<1$ times of the accretion energy
has been re-emitted above polar cap, has been included. It is not
surprising that some of the nearby sources discovered recently
(e.g., CCOs, DTNs, etc.) are actually low-mass {\em bare} strange
stars since their luminosities are generally much smaller than
$\sim 10^{35}$ erg/s.

Crusts forms if one of the two conditions can not be satisfied.
The accretion rate of a star moving in a medium with density
$\rho=\rho_{24}\times 10^{-24}$ g/cm$^3$ is $\sim
10^{-18}M_1^2V_7^{-3}\rho_{24}M_\odot$ yr$^{-1}$, where the
stellar mass $M=M_1M_\odot$, the velocity $V=V_7\times 10^7$ cm/s.
Even for an accretion of Hubble time ($\sim 10^{10}$ years), the
crust mass could be order of $\sim
10^{-8}M_1^2V_7^{-3}\rho_{24}M_\odot$, which is very likely to be
much smaller than the maximum crust mass, $10^{-6\sim -5}
M_\odot$.
The crusts might then be very thin, and the radii of low-mass
strange stars should therefore be very small even for crusted
ones. This is very helpful to identify strange stars.

{\bf 4. Evolutional tracks of pulsar-like stars?}

Besides radio-loud pulsars, radio-quiet pulsar-like compact
objects are also discovered recently in X-ray bands, which include
soft gamma-ray repeaters (SGRs) and anomalous X-ray pulsars (AXPs)
(both have X-ray luminosity $L_{\rm x} \sim 10^{34-35} \,{\rm
erg/s}$), compact center objects (CCOs, $L_{\rm x} \sim 10^{33-34}
\,{\rm erg/s}$), and dim thermal Neutron stars (DTNs, $L_{\rm x}
\sim 10^{32-33}\, {\rm erg/s}$).
Why do pulsar-like stars behave so diversely?

Many factors may affect the star's appearances, whereas we note
that two parameters could be very important for their evolution
scenarios: {\em environmental density} and {\em stellar mass}.
A dense environment may be responsible for leaving a
remnant-nebula after a supernova and for resulting possibly in a
fallback accretion (e.g., via disk) around the compact star.
Various astrophysical manifestations could then be due to
differences of both the stellar masses and the accretion stages.
The supernova remnants with non-pulsars have all been identified
with SN IIL/b, which shows that the center compact objects have
the strongest interaction with the surrounding medium
\citep{Chevalier05}.
Therefore, AXPs/SGRs with remnants and CCOs could be born in dense
environments, but DTNs in less dense ISM.
The compact stars of SGRs/AXPs might be of solar-masses, and their
bursts could be of starquake-induced magnetic-energy release or of
bombardments by strange-planets \citep{xu04a}. Whereas CCOs and
DTNs, without bursts, are of low-masses. DTNs should be very fast
rotators provided that they are born in a sparse ISM.
We may expect that a DTN (e.g., RX J1856) may evolve to a state,
where only Plank-like X-ray is radiated, without any extra
UV-optical emission component, when its circum-stellar material
dissipates by the so-called ``propeller'' mechanism.

{\bf 5. Observational tests for the ``low-mass'' idea}

Several ways, to be summarized, are proposed to test the
``low-mass'' idea.

1, {\em Dust emission around pulsar-like stars}.
Accretion models are suggested to explain millisecond pulsars
(``recycled''), CCOs, DTNs, and AXP/SGRs, but the accretion modes
around these pulsar-like stars is still not yet clear. Anyway, in
case of disk accretion, the central temperature, as a function of
distance to the star, $r$, in the standard model (the
$\alpha$-disk) is $T_c=1.2\alpha^{-1/5}{\dot
M}_{16}^{3/10}M_1^{1/4}r_{10}^{-3/4}$ eV for $r\gg R$ \citep{f92},
where $r_{10}=r/(10^{10}$cm), ${\dot M}_{16}={\dot
M}/(10^{16}$g/s), and the $\alpha$-prescription of viscosity is
taken.
We see then that, outside the light-cylinder, the disk around a
low-mass strange star would have $T_c\sim 0.1$ eV, which emits in
sub-millimeter bands. In addition, normal planets that are still
embedded in young circumstellar disks could also contribute much
submillimeter emission \citep{w04}.
It is thus important to test models \citep{xu04} for pulsar-like
stars by detecting dust emission (with {\em Spitzer} and {\em
SCUBA}, etc.).

2, {\em Determination the radii of distant pulsar-like stars}.
A few ultracompact binaries were discovered (i.e., the orbital
periods, $P_{\rm orb}$, is very short), which include 4U 1543-624
($P_{\rm orb}=18.2$ min), 4U 1820-30 ($P_{\rm orb}=11$ min), RX
J1914.4+2456 ($P_{\rm orb}=569$ s), RX J0806.3+1527 ($P_{\rm
orb}=321$ s). Such ultrashort period binaries are supposed to be
double-degenerated, possibly to be ``NS + WD'' systems.
For instance, the low-mass X-ray binary, 4U 1543-624, has likely a
companion of a C-O white dwarf with mass $\sim 0.03 M_\odot$
\citep{wang04}.
However, an alternative possibility for the nature of these
binaries might be that they could be ``WD + low-mass strange
star'' systems.
This idea could be tested by future X-ray interference telescopes
(e.g., {\em MAXIM}) with an angular resolution of $0.1~\mu$arcsec,
because a low-mass WD with radius $\sim 10^4$ km could be imaged
(having angular size of $\sim 0.7~\mu$arcsec at a distance of 100
pc), but a low-mass strange star (or a strange planet) is still a
point-source even observed by such an advanced telescope.

3, {\em Gravitational wave detection}.
Detection of gravitational waves can certainly test the general
theory of relativity, and also open a new window for us to observe
astrophysical phenomena. Normal neutron stars could be a kind of
gravitational-wave sources, both in the birth stages
\citep{r-mode} and in their later evolutions when glitches occur
\citep{r-mode-glitch}.
However, if pulsar-like stars are actually low mass strange stars,
and could be solidified soon after birth, we should not be able
detect the waves, especially during glitches.
We propose thus to check these issues by {\em LISA} and {\em
LIGO}.

4, {\em Searching for sub-millisecond pulsars}.
Normal neutron stars can {\rm not} spin with frequencies less than
$\sim 0.5M_1^{1/2}R_6^{-3/2}$ ms ($R_6=R/10^6$ cm), but low-mass
bare strange stars can, even less than 0.1 ms \citep{xu04}. A
possibility arises then to answer the question, what the nature of
pulsar-like stars is, through searching for sub-millisecond
pulsars in radio as well as in X-ray (and maybe other) bands.
The shortest period of pulsars (PSR 1937+21) is 1.558 ms, and the
investigation could also probably lead to breaking this record.
We need thus a much short sampling time, and would deal with then
a huge amount of data in order to find a sub-millisecond pulsars.

{\bf 6. Conclusions}

We have studied further low-mass bare strange stars, in addition
to the work of \cite{xu04}.
Low masses strange stars could form via AIC of CO or O-Ne-Mg white
dwarfs if most of the white dwarf masses are ejected during
explosion, being similar to the case of core-collapse supernova of
evolved massive stars.
Ions accreted onto a bare strange star surface may have enough
kinematic energy to penetrate the Coulomb barrier as long as the
stellar mass is higher than $\sim 10^{-4}M_\odot$ and the
luminosity is not much higher than $\sim 10^{34}$ erg/s. Such a
strange star would thus keep to be bare even in an accretion
phase.
A low-mass bare strange star can spin very fast, even to a period
of $<0.1$ ms, due to the color-confinement by itself.
Various astrophysical manifestations of pulsar-like stars may
depend on their circumstellar density, but also on the stellar
masses of residual stars.
Observations to test this ``low-mass'' idea are presented,
including detecting dust emission and gravitational wave,
geometrically determining stellar radii, and searching for
sub-millisecond pulsars.

{\em Acknowledgments}.
This work is supported by National NSF of China (10273001) and the
973 Projects of China (G2000077602). I would like to thank Prof.
Qiao for his stimulating discussions when writing this paper.


\begin{thebibliography}{}

\bibitem[Andersson \& Comer(2001)]{r-mode-glitch}
Andersson, N., Comer, G. L. 2001, Phys. Rev. Lett., 87, 241101

\bibitem[Andersson et al.(2002)]{r-mode}
Andersson, N., Jones, D. I., Kokkotas, K. D. 2002, Mon. Not. Roy.
Astron. Soc. 337, 1224 (astro-ph/0111582)

\bibitem[Canal \& Guti\'errez(1997)]{cg97}
Canal, R., Guti\'errez, J. 1997, in: {\em White, dwarfs}, Proc.
10th Europ. Workshop on White Dwarfs, eds. J. Isern, M. Hernanz,
and E. Gracia-Berro, Kluwer Academic Publishers, p.49
(astro-ph/9701225)

\bibitem[Chevalier(2005)]{Chevalier05}
Chevalier, R. A., 2005, ApJ, in press (astro-ph/0409013)

\bibitem[Frank et al.(1992)]{f92}
Frank, J., King, A., Raine, D. 1992, {\em Accretion power in
astrophysics}, Cambridge Univ. Press (\S5.6).

\bibitem[Fryer et al.(1999)]{f99}
Fryer, C., Benz, W., Herant, M., Colgate, S. A., 1999, ApJ, 516,
892

\bibitem[Wang \& Chakrabarty(2004)]{wang04}
Wang, Z., Chakrabarty D. 2004, ApJ (Letters), in press
(astro-ph/0406465)

\bibitem[Wolf \& D'Angelo(2004)]{w04}
Wolf, S., D'Angelo, G. 2004, ApJ, in press (astro-ph/0410064)

\bibitem[Xu(2002)]{xu02}
Xu, R. X. 2002, ApJ, 570, L65

\bibitem[Xu(2004)]{xu04}
Xu, R. X. 2004, MNRAS, in press (astro-ph/0402659)

\bibitem[Xu(2004a)]{xu04a}
Xu, R. X. 2004a, in: Young neutron stars and their environments,
IAU Symp. 218,  eds. F. Camilo and B. M. Gaensler, p.299

\end{thebibliography}
\end{document}